\documentclass[useAMS,usenatbib]{mn2e}
\usepackage{graphicx}
\usepackage{rotating}
\usepackage{pdflscape}

\title[]
  {Extreme hydrodynamic atmospheric loss near the critical thermal escape regime}
\author[]
  {N. V.~Erkaev,$^{1,2}$, H.~Lammer,$^3$, P. Odert,$^4$, Yu. N. Kulikov,$^5$, K. G.~Kislyakova$^3$
\\
$^1$Institute for Computational Modelling, Russian Academy of Sciences, Krasnoyarsk 36, Russian Federation\\
$^2$Siberian Federal University, Krasnoyarsk, Russian Federation\\
$^3$Space Research Institute, Austrian Academy of Sciences, Schmiedlstr. 6, A-8042, Graz, Austria\\
$^4$Institute of Physics, University of Graz, Universit\"{a}tsplatz 5, A-8010 Graz, Austria\\
$^5$Polar Geophysical Institute, Russian Academy of Sciences, Khalturina Str. 15, Murmansk 183010, Russian Federation}
\date{Released 2014 Xxxxx XX}

\pagerange{\pageref{firstpage}--\pageref{lastpage}} \pubyear{2014}

\def\LaTeX{L\kern-.36em\raise.3ex\hbox{a}\kern-.15em
    T\kern-.1667em\lower.7ex\hbox{E}\kern-.125emX}

\begin{document}

\label{firstpage}

\maketitle

\begin{abstract}
By considering martian-like planetary embryos inside the habitable zone of solar-like stars we study the behavior of the hydrodynamic atmospheric escape of hydrogen for small values of the Jeans escape parameter $\beta$$<$3, near the base of the thermosphere, that is defined as a ratio of the gravitational and thermal energy.
Our study is based on a 1-D hydrodynamic upper atmosphere model that calculates the volume heating rate in a hydrogen dominated thermosphere due to the
absorption of the stellar soft X-ray and extreme ultraviolet (XUV) flux. We find that when the $\beta$ value near the
mesopause/homopause level exceeds a critical value of $\sim$2.5, there exists a steady hydrodynamic solution with a smooth transition from subsonic
to supersonic flow. For a fixed XUV flux, the escape rate of the upper atmosphere is an increasing function of the temperature at the lower boundary.
Our model results indicate a crucial enhancement of the atmospheric escape rate, when the Jeans escape parameter $\beta$ decreases to this critical value. When $\beta$ becomes $\leq$ 2.5, there is no stationary hydrodynamic transition from subsonic
to supersonic flow. This is the case of a fast non-stationary atmospheric expansion that results in extreme thermal
atmospheric escape rates.
\end{abstract}


\section{INTRODUCTION}
Studies related to hydrodynamic escape and evolution of planetary atmospheres
(e.g., Sekiya et al. 1980a, 1980b, 1981; Watson et al. 1981; Kasting \& Pollack 1983; Yelle 2004; Tian et al. 2005a, 2005b; Murry-Clay et al. 2009;
Koskinen et al. 2013; Erkaev et al. 2013, 2014; Lammer et al. 2013a, 2014) indicate that the heating of the upper atmosphere caused by high
XUV radiation and the related hydrodynamic expansion of the bulk atmosphere is important for the escape of light gases such as hydrogen from early planetary atmospheres.
The hydrodynamic outflow of the atmospheric particles is somewhat similar to that of the solar wind described by the well known isothermic model of Parker (1964a, 1964b). For escape, the atmospheric gas requires a heating source
to overcome the gravitational potential of the planetary body. The main heating source in planetary atmospheres is provided by the host star in
the form of XUV radiation and its absorption in the upper atmosphere.

In their pioneering study Watson et al. (1981) proposed an analytical hydrodynamic escape model, where the XUV heating was assumed to be deposited within a thin layer near the planetary surface. In reality the XUV flux is absorbed over a wider altitude range (e.g., Tian et al. 2005a; Murry-Clay et al. 2009). For instance, Murray-Clay et al. (2009) applied a more realistic upper atmosphere heating model
where the XUV volume heating rate $Q_{\rm XUV}$ can be written as
\begin{eqnarray}
Q_{\rm XUV} =  \eta I_{\rm xuv0} exp(-\tau) \sigma n,
\end{eqnarray}
with a heating efficiency $\eta$ that corresponds to the fraction of the absorbed XUV
radiation that is transformed into thermal energy (Shematovich et al. 2014), the XUV absorption cross section $\sigma_{\rm XUV}$
which is $\sim 5\times 10^{-18}$ cm$^{2}$ for a hydrogen dominated atmosphere (Beyont \& Cairns 1965),
the neutral atmosphere density $n$, and optical depth $\tau$
\begin{eqnarray}
\tau = \sigma \int_r^\infty{n dr}.    \label{tau}
\end{eqnarray}
This XUV heating model was further extended by Erkaev et al. (2013) by
taking into account the angular dependence of the heating rate.
Generally, the main necessary condition for hydrodynamic escape modeling is that
the sonic point should be within the collision dominated atmosphere where the Knudsen number $Kn$
is sufficiently small. Volkov et al. (2011) suggested a criterion where hydrodynamic models should
be valid without problems as long as the $Kn\leq$0.1.

In our present study we investigate the escape criteria and efficiency of atomic hydrogen from a Mars-like planetary embryo at 1 AU that experiences XUV flux values between 45--100 times higher than that of today's solar value. A relatively low gravity and temperatures at the base of the thermosphere that are $\geq 250$ K
provide low values of the Jeans escape parameter $\beta$
\begin{equation}
\beta =\frac{GM_{\rm pl}m}{kTR},
\end{equation}
which can cause a crucial enhancement of the hydrodynamic
atmospheric escape rates. With $G$ Newton's gravitational constant, $M_{\rm pl}$ the mass of a planet, $m$ the mass of the atmospheric main species, Boltzmann constant $k$, atmospheric temperature $T$ and planetocentric distance $R$. For large gas giants such as Jupiter the $\beta$ value at the base of the thermosphere is $\sim 474$ and for hot Jupiter's such as HD209458b the
thermosphere base $\beta$ value is $\sim 313$, while at a distance of $\sim 3R_{\rm pl}$ $\beta$ decreased to $\sim 5.6$. For low mass planetary bodies
such as the building blocks of terrestrial planets like martian size planetary embryos that outgassed volatiles which produced steam atmospheres during the solidification of magma oceans (Elkins-Tanton 2012), or nebula captured hydrogen gas (Lammer et al. 2014), that are exposed by the high XUV flux of young host stars inside the habitable zone $\beta$ near the base of the thermosphere can reach values that are $<3$.

With this in mind, the main aim of our study is to investigate the behavior of the hydrodynamic atmospheric expansion of a hydrogen dominated upper atmosphere  for low $\beta$ values. In Sect. 2
we describe the model approach, in Sect. 3 we present and discuss our analytical and numerical results and their relevance for the evolution of initial water inventories of terrestrial planets.
\section{Model equations}
\subsection{Hydrodynamic upper atmosphere model}
To study the XUV-heated upper atmosphere structure and thermal
escape rates of exposed hydrogen atoms, we apply a 1-D hydrodynamic upper atmosphere model with an energy absorption model,
that is described in detail in Erkaev et al. (2013, 2014) and Lammer et al. (2013a, 2014). The
model solves the system of the hydrodynamic equations for mass,
\begin{equation}
\frac{\partial \rho R^2}{\partial t} + \frac{\partial \rho v R^2}{\partial R}= 0,
\end{equation}
momentum,
\begin{eqnarray}
\frac{\partial \rho v R^2}{\partial t} + \frac{\partial \left[ R^2 (\rho v^2+P)\right]}{\partial R} =\rho g R^2 + 2P R,
\end{eqnarray}
and energy conservation,
\begin{eqnarray}
\frac{\partial R^2\left[\frac{\rho v^2}{2}+\frac{P}{(\gamma-1)}\right]}{\partial t}
+\frac{\partial v R^2\left[\frac{\rho v^2}{2}+\frac{\gamma P}{(\gamma - 1)}\right]}{\partial R}=\nonumber\\
\rho v R^2 g + Q_{\rm XUV} R^2.
\end{eqnarray}
The distance $R$ corresponds to the radial distance from the
centre of the planet, $\rho, P, T, v$ are the mass
density, pressure, temperature and velocity of the XUV exposed and heated non-hydrostatic
outward flowing bulk atmosphere. $\gamma$ is the polytropic index,
$g$ the gravitational acceleration and $Q_{\rm XUV}$ is the XUV
volume heating rate (Erkaev et al. 2013, 2014)
\begin{eqnarray}
Q_{\rm XUV}=\eta\sigma_{\rm XUV} n I_{\rm XUV}^* \frac{1}{2}\int_0^{\pi/2+\arccos(1/r)}{J(r,\theta)_{\rm XUV}
\sin(\theta)d\theta},
\end{eqnarray}
where $r$ is the normalized radial distance $r = R/R_0$, $R_0$ is the lower boundary
radius at the base of the thermosphere near the mesopause/homopause level, $\eta$ is the heating efficiency which is typically 15\% (Shematovich et al. 2014),
$I_{\rm XUV}^{*}$ is the XUV flux at the upper boundary, $\sigma_{\rm XUV}$ is the XUV absorption cross section
in a hydrogen atmosphere (Chassefi\`{e}re 1996; Erkaev et al. 2013; Lammer et al. 2014), and $J(r,\theta)_{\rm XUV}$
is a dimensionless function  describing the variation the XUV flux
with respect to the radial distance due to the
atmospheric absorption,
\begin{eqnarray}
J(r,\theta)_{\rm XUV} = exp\left [-\int_r^{R^*}{a\tilde n(\xi)
(\xi^2 - r^2\sin(\theta))^{-1/2}\xi d\xi}\right].  \label{J}
\end{eqnarray}
Here $R^*$ is the upper boundary radius, $\tilde n = n/n_0$, where $n_0$ is the normalized density at the lower boundary, and $a = \sigma_{\rm XUV} n_0 R_0$ is a
dimensionless constant parameter.
Formula (\ref{J}) describes the XUV flux intensity as a function of spherical coordinates $r$ and $\theta$.
In the particular case when $\theta$=0, this formula can be simplified to the expression
\begin{eqnarray}
J(r,\theta)_{\rm XUV}=exp(-\tau),
\end{eqnarray}
where $\tau$ is given by equation (\ref{tau}).
Therefore, our heating model is be quite similar to that of Murray-Clay et al. (2009)
for the central radial direction with zero spherical angle.

In case of a stationary outflow regime, we can neglect the time derivatives and obtain the ordinary system
of equations as follows
\begin{eqnarray}
\rho v R^2 = \Psi , \label{mass} \\
 \rho v\frac{\partial v}{\partial R} + \frac{\partial P}{\partial R} =-\rho\frac{\partial\Phi}{\partial R}   , \label{momentum}\\
\frac{\partial \Psi\left[\frac{v^2}{2}+\frac{\gamma P}{(\gamma - 1)\rho} + \Phi\right]}{\partial R}=
 Q_{\rm XUV} R^2, \label{energy}
\end{eqnarray}
where $\Phi$ is the gravitational potential, and $\Psi$ is the escaping mass flux per unit steradian.

Integrating equation (\ref{energy}) we get a relationship between the escape rate and the total absorbed energy.
\begin{eqnarray}
\Psi = \frac{ \int_1^{R^*}{Q_{\rm XUV} R^2 dR}}{[{v^*}^2 /2 + \Delta\Phi - {v_0}^2 /2 - \gamma/(\gamma-1) (k T_0/m)]}. \label{Escape_rate1}
\end{eqnarray}
Here $v_0$ and $v^*$ are the flow velocities at the lower and upper boundaries.
The total absorbed energy (energy deposition) is related to the incoming XUV flux
\begin{eqnarray}
4\pi\int_1^{R_*}{Q_{\rm XUV} R^2 dR} = \eta  I_{\rm XUV}^* \pi R_{\rm eff}^2,    \label{I}
\end{eqnarray}
where $R_{\rm eff}$ is the effective XUV absorbtion radius
\begin{eqnarray}
R_{\rm eff} = R_0\left [  1 +  {\int_1}^{\tilde R_*}{(1 - \tilde J(s,0) )2 s ds } \right]^{1/2},
\end{eqnarray}
Introducing definition $\beta$ = $m \Delta \Phi /(k T_0) $, and using (\ref{I}) we rewrite equation (\ref{Escape_rate1})
\begin{eqnarray}
4\pi\Psi = \frac{\eta I_{\rm XUV}^* \pi {R_{\rm eff}}^2}{\Delta \Phi} A,  \label{Escape_rate2}\\
 A =\frac{1}{\left[1 + ({v^*}^2 -{v_0}^2) /(2\Delta\Phi) - \gamma/[(\gamma-1) \beta]\right]}.
\end{eqnarray}
For large values of the Jeans escape parameter $\beta$, the effective radius $R_{\rm eff}$ is close to the base of the thermosphere
$R_0$, the ratio ${v^*}^2/\Delta\Phi$ can then be neglected and
the coefficient $A\approx 1$. In such a case equation
(\ref{Escape_rate2})  yields the well known energy limited escape formula
\begin{eqnarray}
4\pi\Psi = \frac{ \eta I_{\rm XUV}^* \pi {R_{\rm eff}}^2}{\Delta \Phi}.
\end{eqnarray}
In cases of a small $\beta$ value near $R_0$, the escape rate becomes much larger
than that predicted by the energy limited escape formula (\ref{Escape_rate1})
because the effective radius ($R_{\rm eff}$)
increases substantially, and in equation (\ref{Escape_rate1}) the terms $k T /m$ and $\Delta\Phi$
become comparable to each other. These terms become equal to each for a critical value of $\beta$
\begin{eqnarray}
m\Delta\Phi / (k T_0) = \beta_{c} = \frac{\gamma}{\gamma-1}. \label{crit}
\end{eqnarray}
In such a case the enthalpy of the lower atmospheric gas is large enough to provide expansion
of the atmosphere against gravity, and an additional heating is not needed for that.
The escape rate is expected to have a pile-up when the Jeans escape parameter $\beta$ approaches to $\beta_c$.
\begin{figure*}
\begin{center}
\includegraphics[width=0.85\columnwidth]{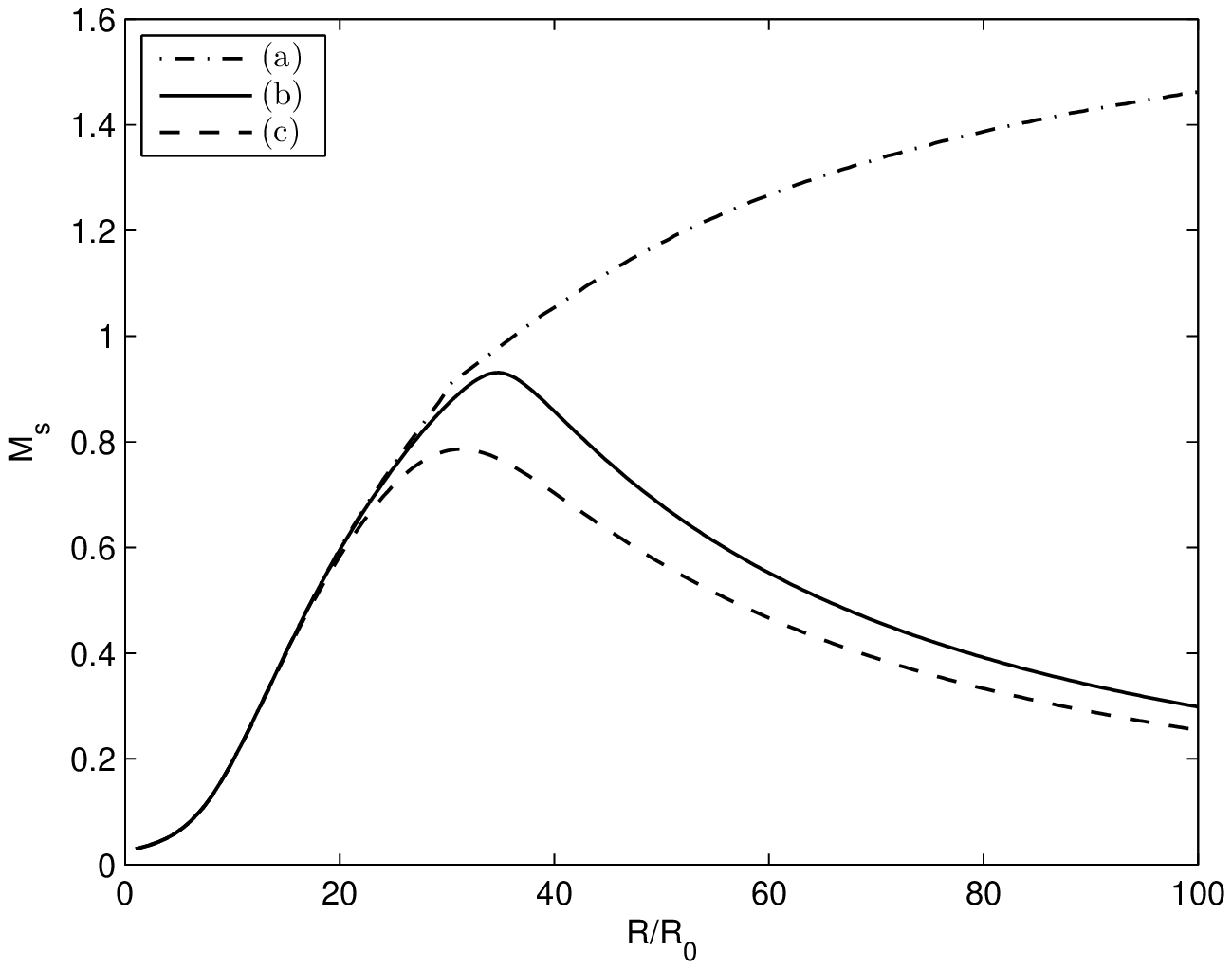}
\includegraphics[width=1.22\columnwidth]{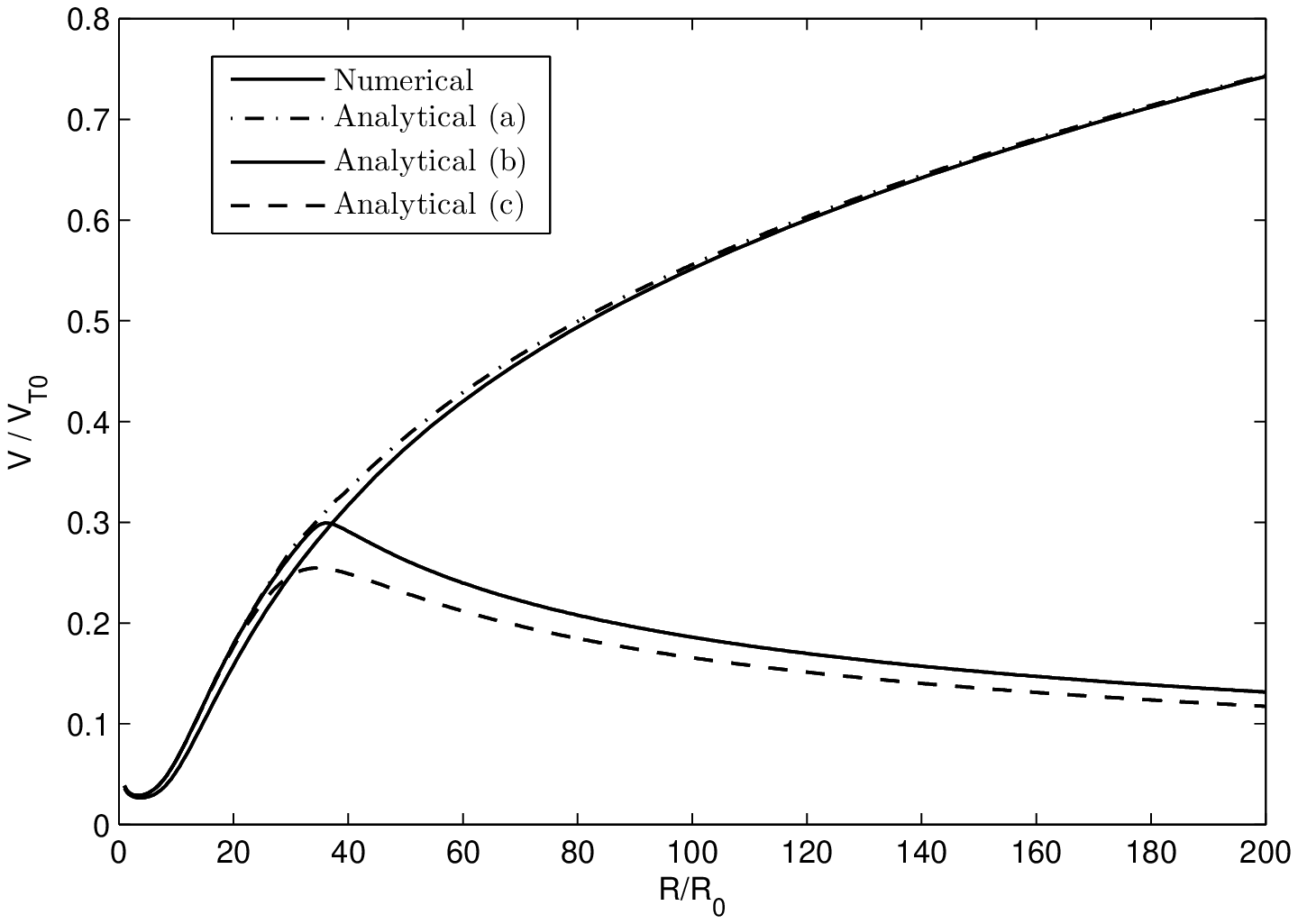}
\caption{Left hand side: Analytical sonic Mach number profiles for $\beta$ =2.7. Right hand side: Analytical and numerical velocity  profiles for $\beta$ =2.7; here curves (a),(b),and (c) correspond to different initial velocity values: 0,03843, 0.03841 and 0.0383, respectively. }
\end{center}
\end{figure*}
\begin{figure*}
\begin{center}
\includegraphics[width=0.95\columnwidth]{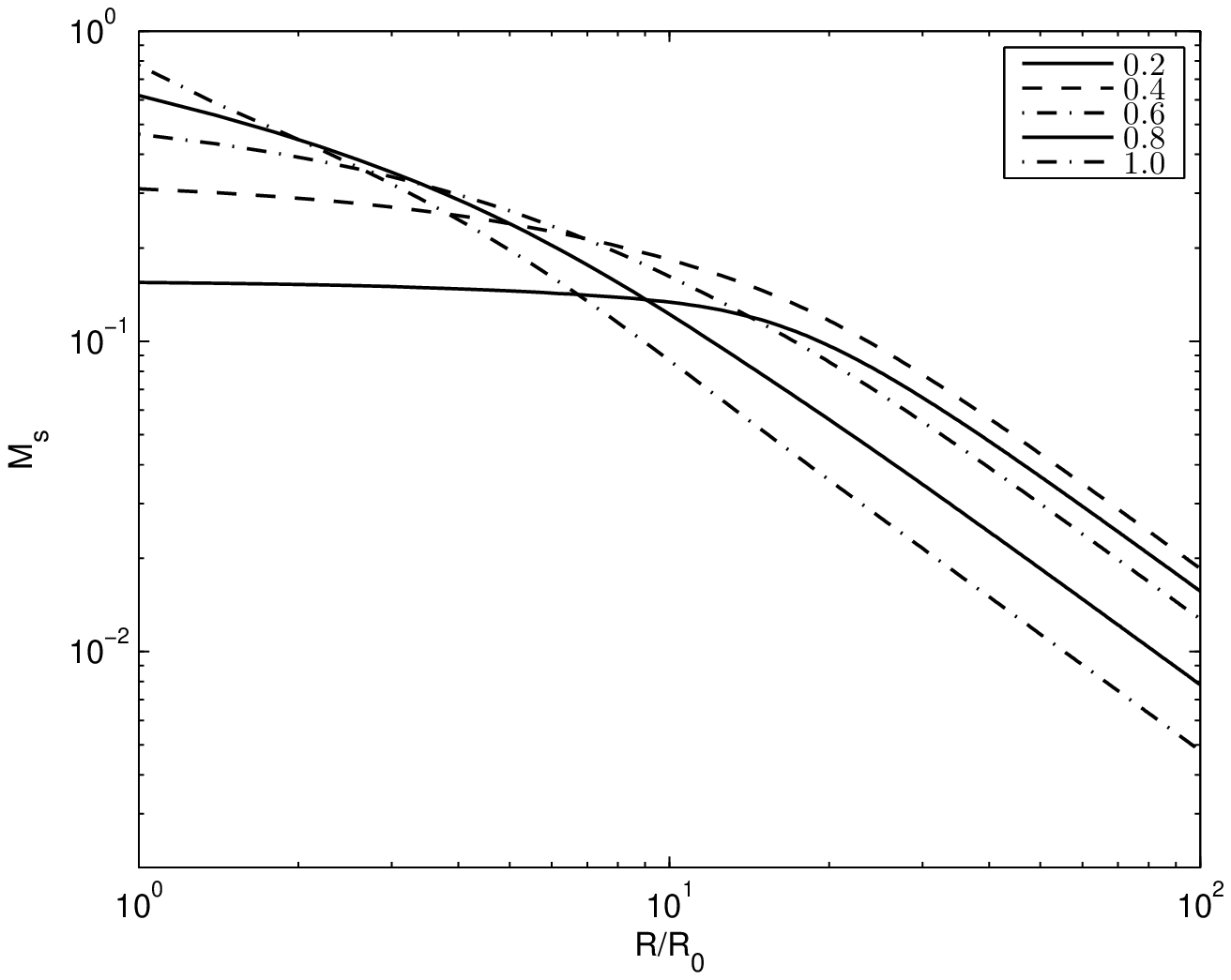}
\includegraphics[width=0.95\columnwidth]{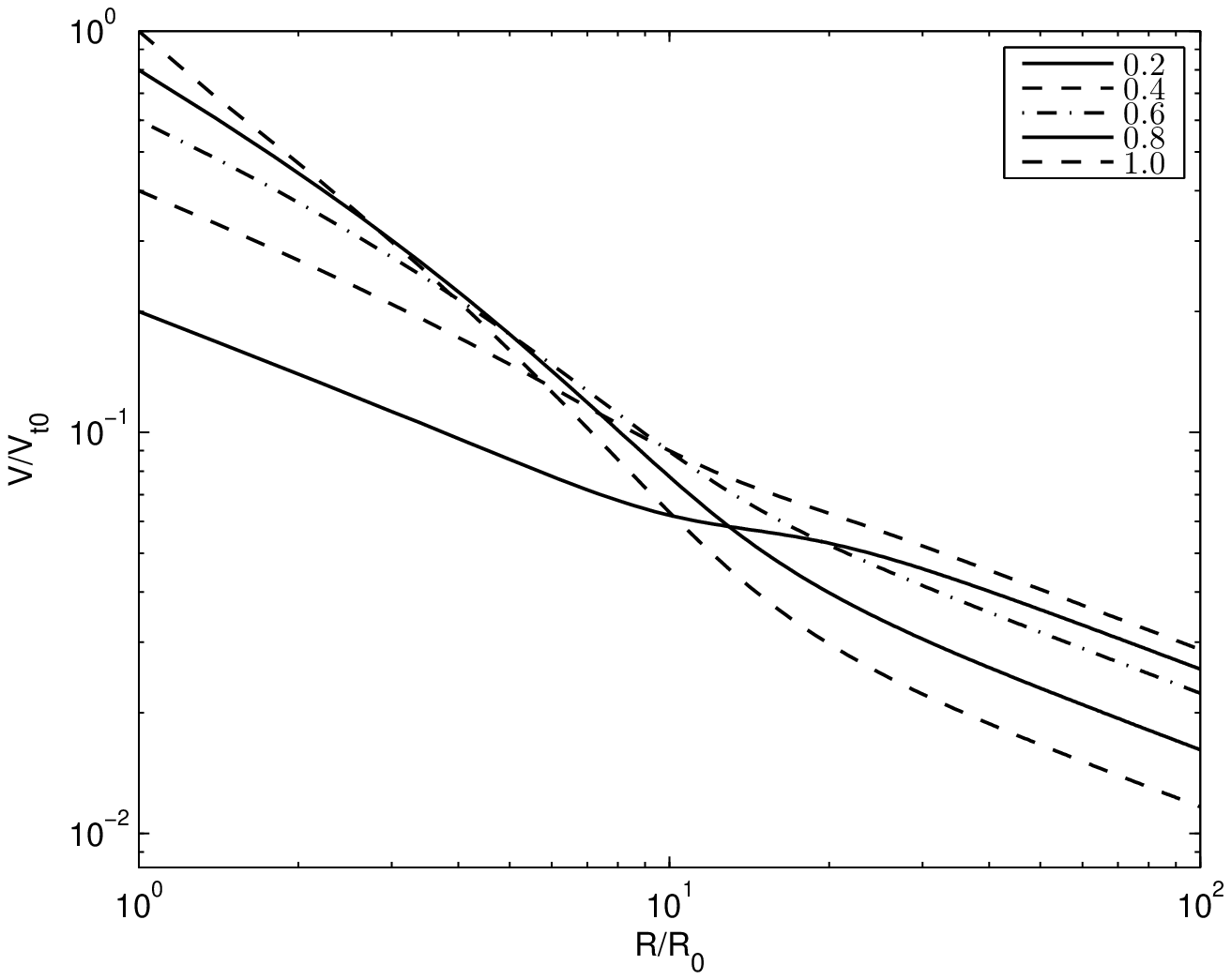}
\caption{Analytical Mach number profiles (left) and velocity profiles (right) corresponding to different initial velocities for $\beta=2.5$}
\end{center}
\end{figure*}


\begin{figure*}
\begin{center}
\includegraphics[width=0.97\columnwidth]{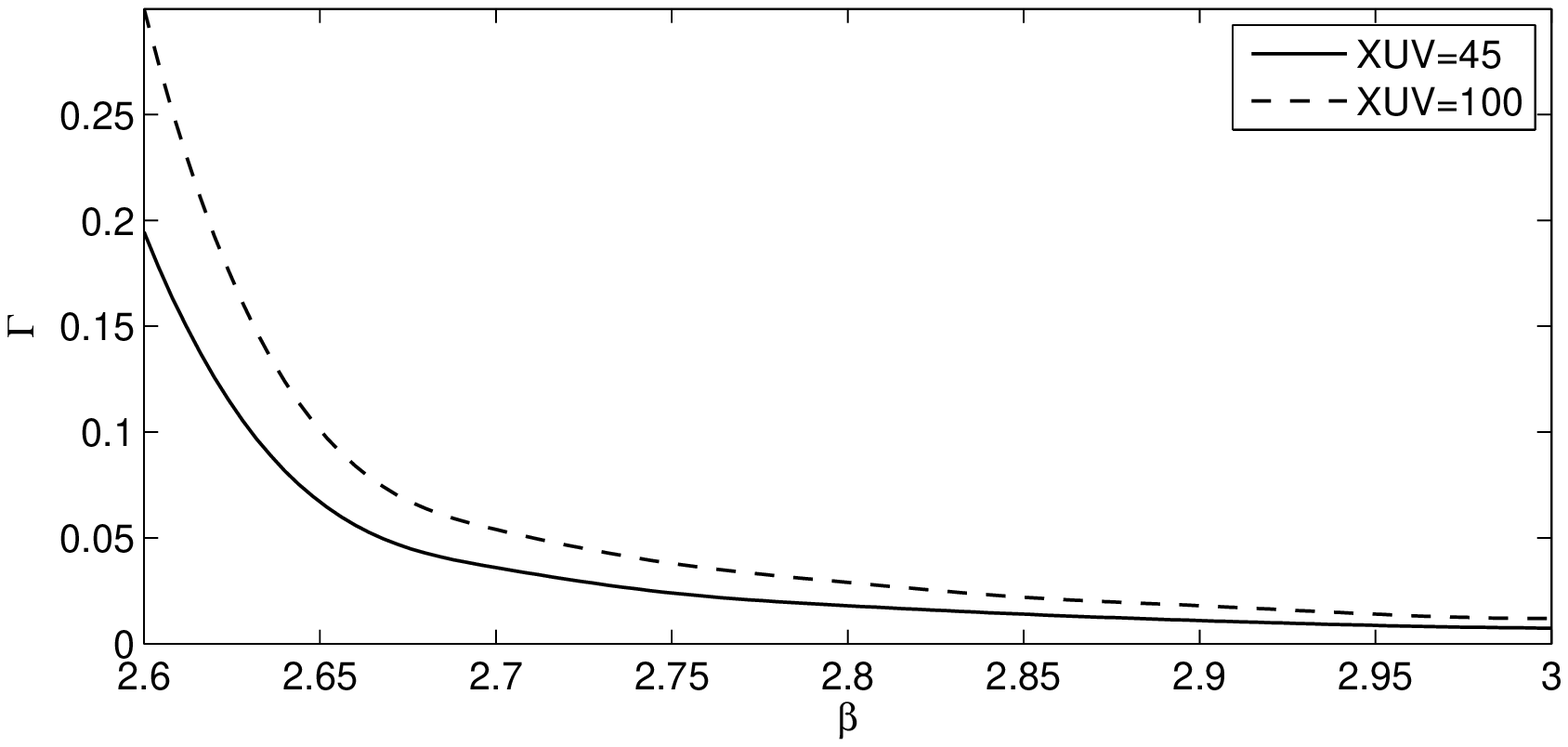}
\includegraphics[width=0.97\columnwidth]{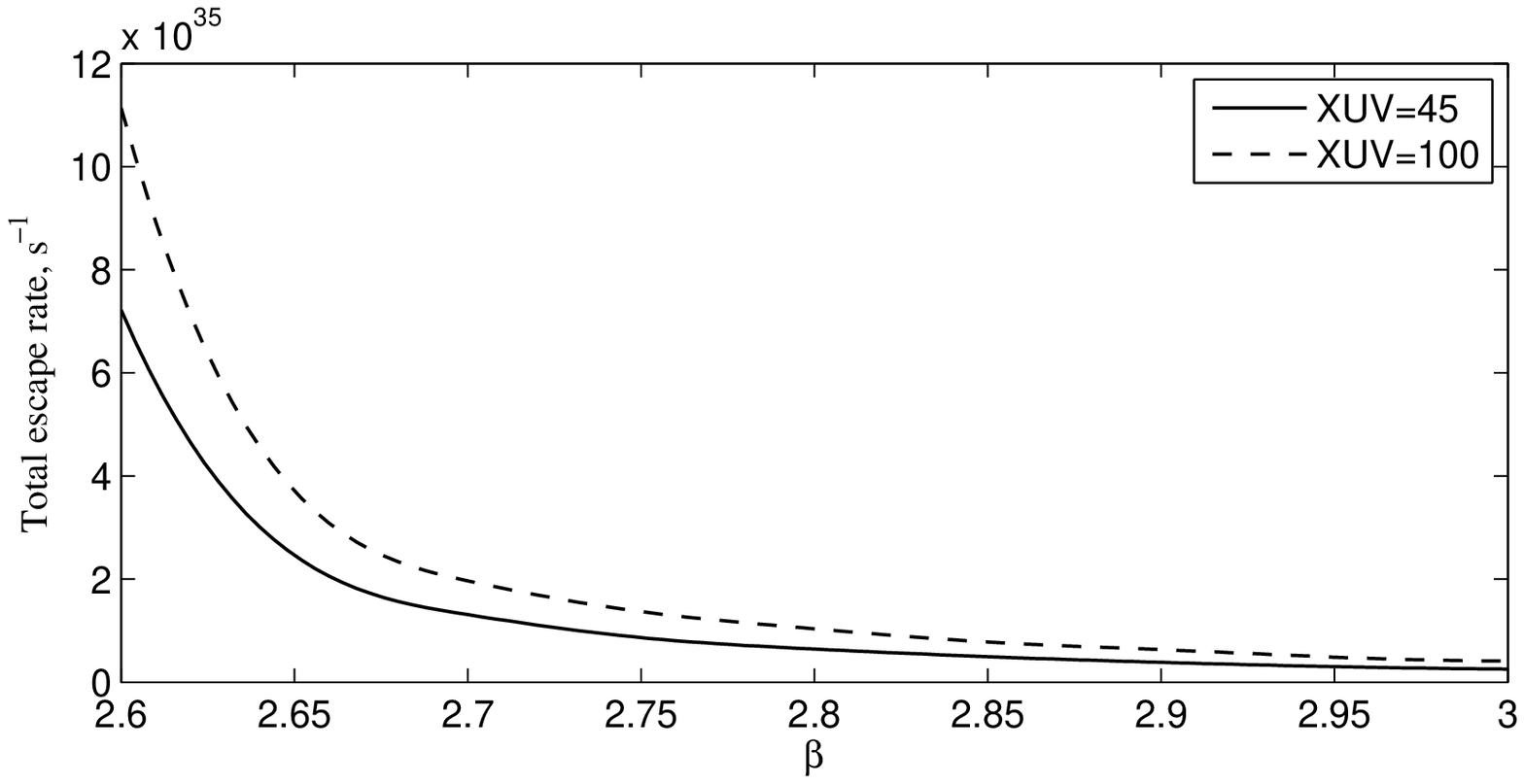}
\caption{Normalized escape rate (per sterad.) and the total escape rate (per sec.) as functions of the Jeans escape parameter $\beta$ for XUV flux values that are 45 and 100 times higher compared to that of today's Sun.}
\end{center}
\end{figure*}
\begin{table*}
\renewcommand{\baselinestretch}{1}
\caption{Main parameters characterizing the hydrodynamic escape flow from a martian-like body at 1 AU, surrounded
by a hydrogen envelope (i.e., dissociated H$_2$O molecules, captured hydrogen from a planetary nebula: $R_{\rm s}$ is the sonic point distance,
and $R^*$ is the distance where the Knudsen number is $\sim$0,1. The $\beta$ values of 2.9, 2.8 and 2.7 correspond to $T_0$ values neat the mesopause/homopause level of 247 K, 256 K, and 266 K, respectively.}
\label{table}
\begin{center}
\begin{tabular}{cccccccccccc}
     XUV & Escape rate, 10$^{33}$ s$^{-1}$ & $R_{\rm eff}/R_0$   & $R_{\rm s}/R_0$  &  $V_0/V_{\rm T_0} $ & $N_{\rm s}/N_0$ & $ T_{\rm s}/T_0$ & $R^*/R_0$ & $V^*/V_{T_0}$ & $N^*/N_0$, 10$^{-9}$ &  $T^*/T_0$ & $\beta$
     \\
    \hline
    \hline
  45& 128  &  13 & 40  &  0,31 & 70  &0,06 & 8000  & 2,6   &  0,29  &1,4  & 2,7\\
  45& 64  & 10  &  40 & 0,31  & 35 & 0,06  & 4500  &  2,1  &  0,48  & 0,9  &   2,8  \\
  45&  40 & 9  &  39 & 0,32  & 34  &  0,06 &3000   &  1,9  &  0,66  &  0,7 &    2,9 \\
\hline
  100&  199 & 13 &31 & 0,37  &  143 & 0,08 &  8000 &  3,5  &  0,25  &  2,3  & 2,7   \\
 100 & 105  & 10 & 31& 0,37  &  81 &  0,08 & 5000   &  2,9  &  0,46  & 1,7   &  2,8 \\
 100 & 63  & 9  & 30  & 0,38  & 54  & 0,08 & 3300  & 2,5   &   0,66 & 1,3   & 2,9 \\
\hline
\end{tabular}
\end{center}
\end{table*}

\section{Analytical and numerical results}
The system of equations (\ref{mass}, \ref{momentum},\ref{energy}) can be transformed to a pair
ordinary differential equations with respect to the normalized entropy and velocity
\begin{eqnarray}
\frac{d S}{d r} = \frac{(\gamma-1){\tilde q} r^{2(\gamma-1)} }{{\tilde V}^{(2-\gamma)} \Gamma^{(\gamma-1)}} , \\ \label{S}
\frac{d \tilde V}{d r} = \frac{{\tilde V}}{r} \frac{\left[{\beta}/r -
2\gamma  S\left( \frac{\Gamma}{{\tilde V}r^2}\right)^{\gamma-1} + (\gamma-1)\frac{r {\tilde q}}{\tilde V}\right]}
{ \left[ \gamma S \left(\frac{\Gamma}{{\tilde V}r^2}\right)^{\gamma-1} -{\tilde V}^2\right]}.\label{V}
\end{eqnarray}
Here $\Gamma$ is the normalized escape rate
\begin{equation}
\Gamma=\frac{\Psi}{(n_0 \sqrt{k T_0 / m} R_0^2)},
\end{equation}
$S$ is the entropy-like function,
$S = {\tilde P}/{\tilde n}^{\gamma-1}$, ${\tilde P}$ is the normalized pressure, $\tilde P = P/(n_0 k T_0))$, ${\tilde V}$ is
the normalized velocity $\tilde V =  V /\sqrt{k T_0/m}$, $r$ is the normalized distance, $r = R/R_0$, and
$\tilde q$ is the normalized heating rate per one particle
\begin{eqnarray}
\tilde q = \frac{Q_{\rm XUV} \sqrt{m} R_0}{n( k T_0)^{3/2} }.
\end{eqnarray}
The denominator of equation (\ref{V}) means a difference between the sonic speed squared and the bulk velocity squared.
At the sonic point, where the denominator vanishes, the numerator must vanish as well. In the subsonic region the
denominator becomes positive, and thus the numerator should be also positive in the most part of the region below the
sonic point in order to provide an acceleration of the flow.
However, near the lower boundary the numerator can be negative in case of a low $\beta$. Therefore, for a low $\beta$ the
numerator has to change its sign in the subsonic region and at the sonic point. In such a case, the velocity behavior
is not monotonic. First it decreases near the lower boundary, and then it starts to increase until the sonic point.
There are two ways for finding the escape rate and radial profiles of the velocity, density and temperature.

The first way is to apply the so-called shooting method for solving the ordinary differential equations with the boundary
condition at the sonic point. This method will allow us to determine the unknown escape rate $\Gamma$. For the particular
unique value of the escape rate it is possible to pass by the sonic point from the subsonic lower boundary to
the supersonic upper boundary.

The second way is to use the time relaxation method by solving the non-steady system of the hydrodynamic equations.
We find the numerical solution by a finite difference numerical scheme described in the previous publications of Erkaev et al. (2013).
Steady-state profiles are obtained by time relaxation of the numerical solution. We used both methods mentioned above and
obtained similar results. In particular we investigate the behavior of the solution when the escape parameter $\beta$
becomes close to the critical value $\gamma/(\gamma-1)$.

The results shown in Figs. 1--3 and Table 1 are based on an assumed hydrogen dominated upper atmosphere that originated either from a hot steam atmosphere
or nebula captured hydrogen gas above a planetary body with the size and mass of Mars in an orbit inside the habitable zone of a Sun-like star.
The high XUV fluxes of young solar-like stars will most likely dissociate hydrogen bearing molecules such as H$_2$O or H$_2$ and
in the upper atmosphere so that it should be mainly dominated by atomic hydrogen (Kastingand \& Pollack 1983; Chassefi\`{e}re 1996;
Yelle 2004; Lammer 2013). During such protoatmosphere conditions one can expect that the surface is much hotter (i.e. solidified magma ocean, magma ocean in mush stage,
surface bombarded frequently by planetesimals) compared to that of present Mars in the Solar System (e.g., Marcq 2013; Lebrun et al. 2013; Erkaev et al. 2014; Lammer et al. 2014). Because of this, the results of 1-D radiative–convective atmospheric models that studied the coupling between hot planetary surfaces and surrounding
steam atmospheres (Marcq 2012) indicate that the mesopause/homopause level $z_0$ of low mass bodies such as Mars may extend from $\sim$100 km as on
present day Mars up to distances that can be located a few 100 Kilometer higher. For this reason we
assume our lower boundary conditions at a distance $R_0=R_{\rm pl}+z_0$ with $z_0=433$ km, with a mesopause/homopause temperature that
correspond to the equilibrium or skin temperature at 1 AU, assumed to be $T_0$=247 K ($\beta$=2.9), 256 K ($\beta$=2.8), and 266 K ($\beta$=2.7). For the number density at that level we assume
a similar value $n_0=5\times 10^{12}$ cm$^{-3}$ (e.g., Tian et al. 2005b; Erkaev et al. 2013, 2014; Lammer et al. 2013, 2014).

On the left hand side, Fig. 1 shows the analytical Mach number profiles corresponding to three different
values of velocity at the lower boundary and a Jeans escape parameter $\beta$ of 2.7. On the right hand side, this figure shows the analytical velocity profiles, and also the numerical profile for $\beta$=2.7. The numerical one is obtained by time relaxation method based on the finite difference
numerical scheme described in Erkaev et al. (2013).
The analytical curves are obtained by the integration of equations (\ref{S}, \ref{V}) for three different initial
velocity conditions: $V_0=0.03843$, 0.03841 and 0.0383.
The true initial value $V_0=0,03843$ corresponds to a separatrix, which provides acceleration of particles
from subsonic to supersonic values. This separatrix is rather close to the profile obtained by the numerical relaxation method.

Fig. 2 shows the analytical sonic Mach number and velocity profiles corresponding to a lower $\beta$ parameter value of 2.5. In this case ($\beta$=2.5),
there is no acceleration at all, when the lower boundary velocity is subsonic. By considering different subsonic velocity conditions,
we obtain only decreasing velocity functions. For any subsonic velocity conditions at the lower boundary, the velocity does not have
monotonic increase until supersonic values. Our results show that the escape rate increases enormously, when the escape parameter $\beta$ approaches
its critical value.

Fig. 3 shows the normalized escape rate (per sterad.) and the total escape rate (per sec.) as functions of the
Jeans escape parameter $\beta$ for XUV flux values that are 45 and 100 times higher compared to that of today's Sun.
For a quite small decrease of the $\beta$ parameter, the escape rate increases by a factor of 10.
However, the sonic point position, and the corresponding velocity and temperature have a minor change
when the escape rate is approaching the critical value. An exception is the density at the sonic point,
which has a substantial increase proportionally to the escape rate.

Table 1 presents the total escape rate, effective radius, and also hydrodynamic parameters
such as the flow velocity, density and temperature corresponding to two particular points.
The first one ($R_{\rm s}$) is the sonic point where the sonic Mach number is $\sim$1.
The second one ($R^*$) is the point where the Knudsen number is $\sim$0.1.
The model results are based of the Jeans escape parameter $\beta$ near the mesopause/hompause level
for three values near the critical one, namely 2.7, 2.8, and 2.9 and two XUV flux intensities
that are 45 and 100 times higher than that of today's solar value.

One can see that with our assumed input parameters, hydrodynamic conditions are
valid far beyond $R_{\rm s}$ is reached. The escape rate rises up to $\sim$3 times
if the $\beta$ value near the critical one, changes only slightly from 2.9 to 2.7. Thus,
extreme uncontrolled blow-off will occur for smaller values.

Chamberlain (1963) and Watson et al. (1981) described $\beta=1.5$ as a critical value for atmospheric ``blow-off'' corresponding to the case
when the thermal energy is equal to the gravitational energy. In the model of Parker (1964a, 1964b), the critical $\beta$ value is equal to 2.
For lower values of $\beta$, a hydrodynamic flow cannot have transition from subsonic to a supersonic regime. A continuous acceleration is
possible only in the supersonic regime. This means that the flow velocity should be supersonic already at the lower boundary.

From our hydrodynamic model we find the critical $\beta$ parameter is $\sim$2.5, which
corresponds to a condition that the enthalpy is equal to the gravitational energy. It is also worth to mention that our results agree with
the kinetic simulations of Volkov et al. (2011) whose simulations yield
a very sharp growth of the escape rate, when the $\beta$ parameter decreases from 2.7 to 2. For $\beta$ values $\leq 2$ this kinetic simulations
predict only supersonic flow. For larger $\beta$ values the kinetic model of Volkov et al. (2011) predicts only subsonic flow
because it does not have heating above $R_0$ because the incoming XUV radiation was assumed to be absorbed around $R_0$.
Or model includes a strong XUV heating above $R_0$, and therefore we have acceleration from a subsonic to a supersonic regime for $\beta > 2.5$.

Small $\beta$ values can be reached on smaller planetary bodies such as
planetary embryos after magma oceans solidified (e.g.,Lammer et al. 2013b; Lebrun et al. 2013),
or lower mass hydrogen-dominated exoplanets in close orbital distances. One can see from Table 1, Mars-type bodies with hydrogen dominated upper atmospheres
will reach this critical conditions quite easy. Smaller bodies will reach the critical hydrodynamic escape regime even easier.
Under extreme conditions the atmosphere at the base of the thermosphere may be hot so that a supersonic flow
can occur around the mesopause/homopause level. Such a very fast non-steady
expansion of an atmosphere may act like an explosion that enhances the
loss of water from the building blocks of the terrestrial planets, that
may finally accrete drier as expected.

H$_2$O and other volatiles may be stepwise
outgassed and potentially lost to a great extent, as illustrated
in Fig. 10 in Lammer et al. (2013b), before accretion ends. If wet planetary embryos lose much of
their initial H$_2$O by fast hydrodynamic escape during their growth to the
final planetary body, the volatile content that is outgassed in
the final stage would be lower than expected. Such a scenario
would agree with the hypothesis of Albar\`{e}de and Blichert-
Toft (2007), which is that the terrestrial planets in the Solar
System accreted dry and obtained most of their water during
the late veneer via impacts.

However, the fast expansion will be connected by strong adiabatic cooling
which has a feedback on the $\beta$ parameter, that will grow above the critical value where a stationary hydrodynamic escape regime
will most likely be established.
\section{Conclusion}
Analyzing the hydrodynamic atmospheric escape solutions for a hot martian-type planetary body (i.e., large planetary embryos
in an inner planetary system, etc.),
we found a critical value for the Jeans escape parameter $\beta_c \sim 2.5$ corresponding to a bifurcation
of the atmospheric expansion regime. A hydrodynamic flow acceleration from subsonic to
supersonic velocities is possible for $\beta$ values that exceed this critical value.
The true value of the escape rate corresponds to the separatrix velocity profile,
which goes from the lower boundary through the sonic point towards infinity.
The escape rate is found to increase crucially, when $\beta$ approaches the critical value.
When $\beta$ is below the critical value, the velocity becomes supersonic even
at the lower boundary. Regarding this, the following scenario is expected.
A supersonic flow at the lower thermosphere means a very fast non-steady expansion of the
atmosphere like an explosion. This expansion might cause a strong adiabatic cooling
of the atmosphere. Such process might result in a decrease of the temperature, and consequently
an increase of the $\beta$ parameter. Finally the $ \beta$ parameter will grow above the critical
value, and thus a stationary hydrodynamic escape can be established.

\section*{ACKNOWLEDGMENTS}
The authors acknowledge the support by the FWF NFN project
S11601-N16 `Pathways to Habitability: From Disks to Active Stars,
Planets and Life', and the related FWF NFN subproject,
S116607-N16 `Particle/Radiative Interactions with
Upper Atmospheres of Planetary Bodies Under Extreme Stellar
Conditions'. P. Odert acknowledges
support from the FWF project P22950-N16. N. V. Erkaev acknowledges
support by the RFBR grant No 12-05-00152-a. Finally, the authors
thank the International Space Science Institute (ISSI) in Bern,
and the ISSI team `Characterizing stellar- and exoplanetary
environments'.

\end{document}